\documentclass[12pt,epsf]{article}
\usepackage{CJKutf8}
\usepackage{amsmath}
\usepackage{booktabs, longtable}
\usepackage{caption2}
\usepackage{cite}
\usepackage{cmap}
\usepackage{enumerate}
\usepackage{fancybox}
\usepackage{flafter}
\usepackage{indentfirst}
\usepackage{latexsym}
\usepackage{listings}
\usepackage{pxfonts}
\usepackage{textcomp}
\usepackage{times}
\usepackage[T1]{fontenc}
\usepackage[dvips]{graphicx}
\usepackage{titletoc}
\usepackage{authblk}
\usepackage{multirow}
\usepackage{subfigure}
\usepackage{hyperref}
\usepackage{color}

\definecolor{lightgray}{rgb}{.7,.7,.7}

\definecolor{red}{rgb}{1,0,0}

\definecolor{blue}{rgb}{0,0,1}

\begin{document}

\begin{center}
{\Large The Universal Property of the Entropy Sum of Black Holes in All Dimensions}
\vskip 1.0cm
Yi-Qiang Du
\footnote{duyiqiang12@mails.ucas.ac.cn}
Yu Tian
\footnote{ytian@ucas.ac.cn}\\
{School of Physics, University of Chinese Academy of Sciences, Beijing 100049, China}
\end{center}
\vskip 1.5cm
\begin{abstract}
It is proposed by Cvetic et al \cite{1} that the product of all horizon areas for general rotating multi-change black holes has universal expressions independent of the mass. When we consider the product of all horizon entropies, however, the mass will be present in some cases, while another new universal property \cite{2} is preserved, which is more general and says that the sum of all horizon entropies depends only on the coupling constants of the theory and the topology of the black hole. The property has been studied in limited dimensions and the generalization in arbitrary dimensions is not straight-forward. In this Letter, we prove a useful formula, which makes it possible to investigate this conjectured universality in arbitrary dimensions for the maximally symmetric black holes in general Lovelock gravity and $f(R)$ gravity. We also propose an approach to compute the entropy sum of general Kerr-(anti-)de-Sitter black holes in arbitrary dimensions. In all these cases, we prove that the entropy sum only depends on the coupling constants and the topology of the black hole.
\end{abstract}

\newpage
{\section{Introduction}}
Studying the black hole entropy has been an attracting work after the establishment of black hole thermodynamics, but it is still a challenge to explain the black hole entropy at the microscopic level. Recently, the microscopic entropy of extreme rotating solutions has drawn some attention, as well as the detailed microscopic origin of the entropy of non-extremal rotating charged black holes. There has been some promising progress and results \cite{5,6}. The further studies of the properties of black hole entropy may give us a deeper understanding of black holes and to study the product of all horizon entropies \cite{1} is an important aspect among them, which is motivated by the following consideration. When the black hole only has an outer horizon and an inner horizon, the inner event horizon plays an important role in studying the black hole physics \cite{chen1,chen2}. For general $4D$ and $5D$ multi-charged rotating black holes, the entropies of the outer and inner horizons are
$$\mathcal{S}_{\pm}=2\pi (\sqrt{N_L}\pm\sqrt{N_R}),$$
respectively, with $N_L$,$N_R$ interpreted as the levels of the left-moving and right-moving excitations of a two-dimensional CFT \cite{a1,a2,a3}. So the entropy product $$\mathcal{S}_+\mathcal{S}_-=4\pi^2(N_L-N_R)$$ should be quantized and must be mass-independent, being expressed solely in terms of quantized angular momenta and other charges. When there are more than two horizons, however, the actual physics of the entropy product or the area product of all the horizons is still not obvious.

Actually, the authors of Ref.\cite{1} have studied the product of all (more than two) horizon areas/entropies for a general rotating multi-charge{d} black hole, both in asymptotically flat and asymptotically anti-de Sitter spacetimes in four and higher dimensions, showing that the area product of the black hole does not depend on its mass $M$, but depends only on its charges $Q_i$ and angular momenta $J_i$. Recently, a new work \cite{4} also studies the entropy product and another entropy relation in the Einstein-Maxwell theory and $f(R)$(-Maxwell) gravity.

As is well-known, in the Einstein gravity (including the theories studied in Ref.\cite{1}), the entropy and the horizon area of the black hole are simply related by $\mathcal{S}=\frac{A}{4}$, so the area product is proportional to the entropy product. However, in (for example) the Gauss-Bonnet gravity where the horizon area and entropy do not satisfy the relation $\mathcal{S}=\frac{A}{4}$ and the entropy seems to have more physical meaning than the horizon area, the mass will be present in the entropy product (see the next section). In fact, Ref. \cite{Giribet} has studied the entropy product by introducing a number of possible higher curvature corrections to the gravitational action, showing that the universality of this property fails in general.

Recently, it is found by Meng et al \cite{2} that the sum of all horizon entropies including ``virtual'' horizons has a universal property that it depends on the coupling constants of the theory and the topology of the black hole, but does not depend on the mass and the conserved charges such as the angular momenta $J_i$ and charges $Q_i$. The conjectural property has only been discussed in limited dimensions. It is believed that the property of entropy sum is more general than that of the entropy product. In this Letter, we prove a useful formula that makes it possible for us to investigate the universal property in all dimensions. Based on this formula, we discuss the entropy sum of general maximally symmetric black holes in the Lovelock gravity, $f(R)$ gravity. As well, we propose a method to calculate the entropy sum of Kerr-(anti-)de-Sitter (Kerr-(A)dS) black holes in the Einstein gravity. In all these cases, we prove that the entropy sum depends only on the coupling constants of the theory and the topology of the black holes. Note that here we just focus on the universal properties, and the actual physics behind it still needs to be further investigated.

This Letter is organized as follows. In the next section, we will discuss the Gauss-Bonnet case, and then we will express the formula and give a brief proof. In the sections 4 and 5, we will use the formula to calculate the entropy sum of (A)dS black holes in the Einstein-Maxwell theory and the Lovelock gravity in all dimensions. In the section 6, we will study rotating black holes to calculate the entropy sum of Kerr-(A)dS metrics in arbitrary dimensions. In the section 7, we will discuss the $f(R)$ gravity where the universal property also holds. At last, we give the conclusion and brief discussion.

{\section{(A)dS black holes in the Gauss-Bonnet gravity}}
The action of the Einstein-Gauss-Bonnet-Maxwell in $d$ dimensions is
\begin{equation}
  I=\frac{1}{16\pi G}\int d^dx\sqrt{-g}[R-2\Lambda+\alpha(R_{\mu\nu\kappa\lambda}R^{\mu\nu\kappa\lambda}-4R_{\mu\nu}R^{\mu\nu}+R^2)-F_{\mu\nu}F^{\mu\nu}]
\end{equation}
Here $G$ is the Newton constant in $d$ dimensions, $\alpha$ is the Gauss-Bonnet coupling constant, and $\Lambda=\pm\frac{(d-1)(d-2)}{2l^2}$ is the cosmological constant. Varying this action with respect to the metric tensor gives equations of motion, which admits the $d$-dimensional static charged Gauss-Bonnet-(A)dS black hole solution \cite{162,163,7,8,9}
\begin{equation}\label{e14}
  ds^2=-V(r)dt^2+\frac{dr^2}{V(r)}+r^2d\Omega_{d-2}^2
\end{equation}
where $d\Omega_{d-2}^2$ represents the line element of a $(d-2)$-dimensional maximal symmetric Einstein space with constant curvature $(d-2)(d-3)k$, and $k=-1, 0$ and $1$, corresponding to the hyperbolic, planar and spherical topology of the black hole horizon, respectively. The function $V(r)$ in the metric \eqref{e14} is given by
\begin{eqnarray}\label{e13}
V(r)=k+\frac{r^2}{2\tilde\alpha}(1-\sqrt{1+\frac{64\pi\tilde\alpha M}{(d-2)r^{d-1}}-\frac{2\tilde\alpha Q^2}{(d-2)(d-3)r^{2d-4}}+\frac{8\tilde\alpha\Lambda}{(d-1)(d-2)}}),
\end{eqnarray}
where $\tilde\alpha=(d-3)(d-4)\alpha$, $M$ and $Q$ are the black hole mass and black hole charge respectively. Horizons of the black holes are located at the roots of $V(r)=0$. The entropy is
\begin{equation}
  \mathcal{S}=\frac{\Omega_{d-2}r^{d-2}}{4}(1+\frac{2(d-2)k\tilde\alpha}{(d-4)r^2}),
\end{equation}
where $\Omega_{d-2}=2\pi^{(d-1)/2}/{\Gamma(\frac{d-1}{2})}$. The area of the horizon is
\begin{equation}
  A=\frac{\Omega_{d-2}r^{d-2}}{4}.
\end{equation}

When we consider the five dimensional charged black hole, according to the function \eqref{e13}, the equation that determines the horizons is 
\begin{equation}\label{e15}
  2\Lambda r^6-12kr^4+(64\pi M-12k^2\tilde\alpha)r^2-Q^2=0.
\end{equation}
Then, we can calculate the product of the areas by using Vieta's theorem and \eqref{e15}
\begin{equation}
  \displaystyle\prod_{i=1}^6A_i=(\frac{\Omega_3}{4})^6\displaystyle\prod_{i=1}^6r_i^3=(\frac{\Omega_3}{4})^6(\frac{-Q^2}{2\Lambda})^3.
\end{equation}
The result does not include the mass $M$, preserving the property revealed in Ref.\cite{1}.

As we have mentioned in the Introduction, the entropy seems to have more physical meaning than the horizon area in the case that the horizon area and entropy are not proportional to each other. In five dimensions, the entropy product has been calculated  when $\Lambda=0$ \cite{Giribet}. Here we will give the explicit result with a non-vanishing cosmological constant $\Lambda$. The product of the entropies is
\begin{equation}
  \displaystyle\prod_{i=1}^6\mathcal{S}_i=(\frac{\Omega_3}{4})^6\displaystyle\prod_{i=1}^6(r_i^3+k\tilde\alpha r_i)=-(\frac{\Omega_3}{4})^6\frac{Q^2}{4\Lambda^2}[Q^2+(64\pi M-12k^2\tilde\alpha)k\tilde\alpha+12k^3\tilde\alpha^2+2\Lambda k^3\tilde\alpha^3]
\end{equation}
and the result depends on the mass.

However, it seems that the sum of all entropies including non-physical entropies proposed by \cite{2} has a better performance, which depends only on the coupling constants of the theory and the topology of the black holes. We find that the Gauss-Bonnet case, which is included in the Lovelock gravity, obeys the property in all dimensions, and we will give the proof later.

{\section{A useful formula}}
In this section, we will prove a formula, which is useful in the following sections.
With regard to the polynomial as follows:$$ a_mr^m+a_{m-1}r^{m-1}+\dots+a_0r^0=0,$$
we denote the roots as $r_i, i=1, 2\cdots m$, and denote $s_n=\displaystyle\sum_{i=1}^mr_i^n$, then we have
\begin{eqnarray}\label{e5}
s_n=\frac{-1}{a_m}\displaystyle\sum_{i=0}^{m-1}s_{n-m+i}a_i,
\end{eqnarray}
 with $s_{n-m+i}=0$ for $n-m+i<0$ and $s_{n-m+i}=n$ for $n-m+i=0$.

The proof is  briefly described as follows:
\begin{eqnarray*}
\begin{split}
\frac{-1}{a_m}(a_{m-1}s_{n-1}+a_{m-2}s_{n-2})=(r_1^n+\cdots+r_m^n)-\displaystyle\sum_{i=1}^m [r_i^{n-2}(\displaystyle\sum_{0<j_1<j_2<m+1,j_1,j_2\neq i}r_{j_1}r_{j_2})],
\end{split}
\end{eqnarray*}
\begin{eqnarray*}
\frac{-1}{a_m}(a_{m-1}s_{n-1}+a_{m-2}s_{n-2}+a_{m-3}s_{n-3})=(r_1^n+\cdots+r_m^n)+\displaystyle\sum_{i=1}^m [r_i^{n-3}(\displaystyle\sum_{0<j_1<j_2<j_3<m+1,j_1,j_2,j_3\neq i}r_{j_1}r_{j_2}r_{j_3})].
\end{eqnarray*}
Continue the process,
if $m\ge n$,
\begin{eqnarray*}
\begin{split}
\frac{-1}{a_m}(a_{m-1}s_{n-1}+a_{m-2}s_{n-2}+\cdots+a_{m-n+1}s_{1})
=(r_1^n+\cdots+r_m^n)+(-1)^nn\displaystyle\sum_{0<j_1<\cdots<j_{n}<m+1}r_{j_1}\cdots r_{j_{n}},
\end{split}
\end{eqnarray*}
so if we set $s_0=n$, then
\begin{eqnarray*}
\begin{split}
\frac{-1}{a_m}&(a_{m-1}s_{n-1}+a_{m-2}s_{n-2}+\cdots+a_{m-n+1}s_{1}+a_{m-n}s_{0})
=r_1^n+\cdots+r_m^n.
\end{split}
\end{eqnarray*}

If $m<n$, we continue the process until $a_{m-l}=a_0$, with $1\leq l\leq m$, one can also find that
\begin{eqnarray*}
\frac{-1}{a_m}\displaystyle\sum_{i=0}^{m-1}s_{n-m+i}a_i=r_1^n+\cdots+r_m^n.
\end{eqnarray*}

{\section{(A)dS black holes in the Einstein-Maxwell theory}}
The Einstein-Maxwell action in $d$ dimensions is
\begin{equation}
  I=\frac{1}{16\pi G}\int d^dx\sqrt{-g}[R-F_{\mu\nu}F^{\mu\nu}-2\Lambda].
\end{equation}

In the maximally symmetric case, solving the equation of motion from the above action gives the RN-(A)dS solution, which is of the form \eqref{e14}.
The horizons are located at the roots of the function $V(r)$ \cite{120,121,122,12}
\begin{eqnarray}\label{e8}
V(r)=k-\frac{2M}{r^{d-3}}+\frac{Q^2}{r^{2(d-3)}}-\frac{2\Lambda}{(d-1)(d-2)}r^2.
\end{eqnarray}
The entropy of horizon is given by
\begin{eqnarray}
\mathcal{S}_i=\frac{A_i}{4}=\frac{\pi^{(d-1)/2}}{2\Gamma(\frac{d-1}{2})}r_i^{d-2}.
\end{eqnarray}

In odd dimensions,just as \cite{2} has showed, the radial metric function is a function of $r^2$ and the entropy $\mathcal{S}_i$ is a function of $r_i$ with odd power. The pairs of roots $r_i$ and $-r_i$ vanish the entropy sum, i.e. $\sum_i\mathcal{S}_i=0$.

In even dimensions, according to equations \eqref{e5} and \eqref{e8}, we have
\begin{equation*}
\begin{split}
  s_{d-2}=\displaystyle\sum_{i=1}^{2(d-2)}r_i^{d-2}&=\frac{-a_{2d-6}}{a_{2d-4}}s_{d-4}=\cdots=(\frac{-a_{2d-6}}{a_{2d-4}})^{\frac{d-4}{2}}s_2\nonumber\\
  &=2(\frac{-a_{2d-6}}{a_{2d-4}})^{\frac{d-2}{2}}=2(\frac{(d-1)(d-2)k}{2\Lambda})^{(d-2)/2}.
\end{split}
\end{equation*}
Then we get
\begin{equation}
  \sum_i\mathcal{S}_i=\sum_i\frac{A_i}{4}=\frac{\pi^{(d-1)/2}}{\Gamma(\frac{d-1}{2})}(\frac{(d-1)(d-2)k}{2\Lambda})^{(d-2)/2}
\end{equation}
which depends only on the cosmological constant $\Lambda$ and the horizon topology k.

To summarize briefly, considering all the horizons including the un-physical ``virtual'' horizons, we find out the general expression of the entropy sum, which depends only on the cosmological constant and the topology of the horizon.

{\section{Black holes in the Lovelock gravity}}
In this section, we will discuss the case of Lovelock gravity. The action of general Lovelock gravity can be written as\cite{150,15}
\begin{equation}
  I=\int d^dx(\frac{\sqrt{-g}}{16\pi G}\displaystyle\sum_{k=0}^m\alpha_kL_k+\mathcal{L}_{matt})
\end{equation}
with $\alpha_k$ the coupling constants and
\begin{equation}
  L_k=2^{-k}\delta_{c_1d_1\cdots c_kd_k}^{a_1b_1\cdots a_kb_k}R^{c_1d_1}_{a_1b_1}\cdots R^{c_kd_k}_{a_kb_k},
\end{equation}
where $\delta^{ab\cdots cd}_{ef\cdots gh}$ is the generalized delta symbol which is totally antisymmetric in both sets of indices.
If only keeping $\alpha_0=-2\Lambda$ and $\alpha_1=1$ nonvanishing, we obtain the Einstein gravity, while keeping $\alpha_2$ nonvanishing as well, we get the Gauss-Bonnet gravity.

Varying the above action with respect to the metric tensor and then solving the resultant equation of motion \cite{160,161,162,163,164,17,18,19} by assuming that the metric has the form \eqref{e14}, one can find that the function $V(r)$ is determined  by
\begin{equation}
  \frac{d-2}{16\pi}\Omega_{d-2}r^{d-1}\displaystyle\sum_{k=0}^N\tilde\alpha_k(\frac{1-V(r)}{r^2})^k-M+\frac{Q^2(d-2)\Omega_{d-2}}{16\pi r^{d-3}}=0,
\end{equation}
where $$N=[\frac{d}{2}],\quad\tilde\alpha_0=\frac{\alpha_0}{(d-1)(d-2)},\quad\tilde\alpha_1=\alpha_1, \quad\tilde\alpha_{k>1}=\alpha_k\displaystyle\prod_{j=3}^{2k}(d-j).$$
This is a polynomial equation for $V(r)$ with arbitrary degree $N$, so generically there is no explicit form of solutions. However, assuming $V(r)=0$ in the above equation, we can also find that horizons of the black holes are located at the roots of the following equation
\begin{equation}\label{e9}
  \frac{d-2}{16\pi}\Omega_{d-2}r^{2d-4}\displaystyle\sum_{k=0}^N\tilde\alpha_k(\frac{1}{r^2})^k-Mr^{d-3}+\frac{Q^2(d-2)\Omega_{d-2}}{16\pi}=0,
\end{equation}
The entropy of horizon is given by
\begin{equation}\label{e10}
  \mathcal{S}=\frac{d-2}{4}\Omega_{d-2}r^{d-2}\displaystyle\sum_{k=1}^N\frac{\tilde\alpha_kk}{d-2k}(\frac{1}{r^2})^{k-1}.
\end{equation}

In odd dimensions, $\sum_i\mathcal{S}_i=0$ with the same reason as before.

For the even dimensions, according to \eqref{e5} and \eqref{e9}, when we calculate $\displaystyle\sum_{j=1}^{2d-4}r_j^{d-2}$,
$$s_{d-2}=\displaystyle\sum_{j=1}^{2d-4}r_j^{d-2}=\frac{-a_{2d-5}}{a_{2d-4}}s_{d-3}+\cdots+\frac{-a_{d-2}}{a_{2d-4}}s_0,$$
we only use the coefficient of $r$ whose power is not smaller than $d-2$, so the mass $M$ and the charge $Q$ will not be present for they belong to the coefficients $a_{d-3}$ and $a_0$ respectively. When we calculate the sum of the entropy \eqref{e10}, the sum of the highest power of roots is $\displaystyle\sum_{j=1}^{2d-4}r_j^{d-2}$, so the mass $M$ and the charge $Q$ will be disappear in the sum of the other power of roots according to \eqref{e5}. It is suggested that the sum of the entropies is independent of mass and charge, just depends on the coupling constants of the theory and the topology constants of the horizon.

{\section{Kerr-(anti-)de-Sitter black holes}}
 Thus far we have only considered the maximally symmetric black holes. It is of great interest to investigate the entropy sum of rotating black holes, albeit in the Einstein gravity. In this section, we will discuss the sum of the entropies in Kerr-de Sitter metrics of all dimensions \cite{200,201,20,21,22}. It is necessary to deal with the case of odd dimensions and that of even dimensions separately.
{\subsection{odd dimensions}}
In odd spacetime dimensions, $d=2n+1$, the equation that determines the horizons can be written as
\begin{equation}\label{e3}
\frac{1}{r^2}(1-\Lambda r^2)\displaystyle\prod_{i=1}^n(r^2+a_i^2)-2M=0
\end{equation}
where $\Lambda$ is the cosmological constant.
The area of the horizon is given by
\begin{equation}\label{e4}
  A_j=\frac{\mathcal{A}_{2n-1}}{r_j}\displaystyle\prod_{i=1}^n\frac{r^2_j+a_i^2}{1+\Lambda a_i^2}
\end{equation}
where
\begin{equation}
  \mathcal{A}_m=\frac{2\pi^{(m+1)/2}}{\Gamma[(m+1)/2]}.
\end{equation}
The entropy is $\mathcal{S}_i=\frac{A_i}{4}$. The sum of the area \eqref{e4} can be
divided into two parts:$$\displaystyle\sum_{j=1}^{2n+2}[A_j-\frac{\mathcal{A}_{2n-1}}{r_j}\displaystyle\prod_{i=1}^{n}\frac{a_i^2}{1+\Lambda a_i^2}]
~~\mbox{and}~~\displaystyle\sum_{j=1}^{2n+2}[\frac{\mathcal{A}_{2n-1}}{r_j}\displaystyle\prod_{i=1}^{n}\frac{a_i^2}{1+\Lambda a_i^2}].$$

The first part is a function of $r$ with odd power. The horizon function \eqref{e3} is a function of $r^2$, which results in roots $r_i$ and $-r_i$ in pair and vanishes the first part. In the second part,$$\displaystyle\sum_{j=1}^{2n+2}[\frac{\mathcal{A}_{2n-1}}{r_j}\displaystyle\prod_{i=1}^{n}\frac{a_i^2}{1+\Lambda a_i^2}]=\mathcal{A}_{2n-1}\displaystyle\prod_{i=1}^{n}\frac{a_i^2}{1+\Lambda a_i^2}\frac{\displaystyle\sum_{0<i_1<i_2<\dots<i_{2n+1}<2n+3}r_{i_1}r_{i_2}\dots r_{i_{2n+1}}}{r_1r_2\dots r_{2n+2}},$$
so it also vanishes because we can find $\displaystyle\sum_{0<i_1<i_2<\dots<i_{2n+1}<2n+3}r_{i_1}r_{i_2}\dots r_{i_{2n+1}}$ vanishes from \eqref{e3} according to Vieta's theorem. Therefore, the sum of entropies vanishes, i.e. $\sum_i\mathcal{S}_i=0$.
{\subsection{even dimensions}}
In even dimensions, $d=2n$, the equation that determines the horizons can be written as
\begin{equation}\label{e6}
\frac{1}{r}(1-\Lambda r^2)\displaystyle\prod_{i=1}^{n-1}(r^2+a_i^2)-2M=0.
\end{equation}

The area of the horizon is given by
\begin{equation}\label{e7}
  A_j=\mathcal{A}_{2n-2}\displaystyle\prod_{i=1}^{n-1}\frac{r^2_j+a_i^2}{1+\Lambda a_i^2}.
\end{equation}
The sum of all the areas (\ref{e7}) is difficult to calculate directly.
However, we can calculate it by the following trick. By using (\ref{e6}), the sum can be recast as
\begin{equation}\label{sum}
\displaystyle\sum_{j=1}^{2n}A_j=\frac{\mathcal{A}_{2n-2}}{\displaystyle\prod_{i=1}^{n-1}(1+\Lambda a_i^2)}\displaystyle\sum_{j=1}^{2n}\frac{2Mr_j}{1-\Lambda r_j^2}=\frac{\mathcal{A}_{2n-2}M}{\sqrt{\Lambda}\displaystyle\prod_{i=1}^{n-1}(1+\Lambda a_i^2)}\displaystyle\sum_{j=1}^{2n}[\frac{1}{1-\sqrt{\Lambda}r_j}-\frac{1}{1+\sqrt{\Lambda}r_j}].
\end{equation}
Firstly, we focus our attention on the $$\displaystyle\sum_{j=1}^{2n}\frac{1}{1-\sqrt{\Lambda}r_j}$$
term in the right hand side of (\ref{sum}). Let $1-\sqrt{\Lambda}r=:\tilde r.$
Then, by substituting $\tilde r$ for $r$, \eqref{e6} develops into
\begin{equation}
(2\tilde r-{\tilde r}^2)\frac{1}{\Lambda^{n-1}}\displaystyle\prod_{i=1}^{n-1}({\tilde r}^2-2\tilde r+1+a_i^2\Lambda)+\frac{2M\tilde r}{\sqrt{\Lambda}}-\frac{2M}{\sqrt{\Lambda}}=0.
\end{equation}
The coefficient of $\tilde r$ in the above equation is $$a_1=\frac{2}{\Lambda^{n-1}}\displaystyle\prod_{i=1}^{n-1}(1+a_i^2\Lambda)+\frac{2M}{\sqrt{\Lambda}},$$ and the constant term of the equation is $$a_0=\frac{-2M}{\sqrt{\Lambda}}.$$ So we obtain
\begin{equation}
 \displaystyle\sum_{j=1}^{2n}\frac{1}{1-\sqrt{\Lambda}r_j}=\displaystyle\sum_{j=1}^{2n}\frac{1}{\tilde r_j}=-\frac{a_1}{a_0}=\frac{\sqrt{\Lambda}}{M\Lambda^{n-1}}\displaystyle\prod_{i=1}^{n-1}(1+a_i^2\Lambda)+1.
\end{equation}
Similarly, we can get
\begin{equation}
 \displaystyle\sum_{j=1}^{2n}\frac{1}{1+\sqrt{\Lambda}r_j}=-\frac{\sqrt{\Lambda}}{M\Lambda^{n-1}}\displaystyle\prod_{i=1}^{n-1}(1+a_i^2\Lambda)+1.
\end{equation}
Therefore the sum of entropies is
$$\displaystyle\sum_{j=1}^{2n}\mathcal{S}_j=\frac 1 4\displaystyle\sum_{j=1}^{2n}A_j=\frac{\mathcal{A}_{2n-2}}{2\Lambda^{n-1}},$$
which depends only on $\Lambda$. The result is independent of the signature of $\Lambda$.

{\section{(A)dS black holes in the $f(R)$ gravity}}
In this section, we consider the action of $R+f(R)$ gravity coupled to a Maxwell field in d-dimensional spacetime\cite{30,31,3}
\begin{equation}
  I=\int d^dx\sqrt{-g}[R+f(R)-(F_{\mu\nu}F^{\mu\nu})^p]
\end{equation}
where $f(R)$ is an arbitrary function of scalar curvature $R$. Solving the corresponding equation of motion in the maximally symmetric case again gives a solution of the form \eqref{e14}, where the function $V(r)$ is given by
\begin{eqnarray}\label{e12}
V(r)=k-\frac{2M}{r^{d-3}}+\frac{Q^2}{r^{d-2}}\frac{(-2Q^2)^{(d-4)/4}}{1+f^{'}(R_0)}-\frac{2\Lambda_f}{(d-1)(d-2)}r^2
\end{eqnarray}
with $f^{'}(R_0)=\frac{\partial f(R)}{\partial R}\mid_{R=R_0}, R_0=\frac{2d}{d-2}\Lambda_f$, $\Lambda_f$ is the cosmological constant. $V(r)=0$ gives the horizons of the black holes.

The entropy of horizon is given by
\begin{eqnarray}
\mathcal{S}_i=\frac{A_i}{4}(1+f^{'}(R_0)),
\end{eqnarray}
and the area of the horizon is given by
\begin{equation}
A_i=\frac{2\pi^{(d-1)/2}}{\Gamma(\frac{d-1}{2})}r_i^{d-2}.
\end{equation}

According to equations \eqref{e5} and \eqref{e12}, in odd dimensions, considering $s_1=\displaystyle\sum_{i=1}^dr_i=0$, we obtain
\begin{equation}
  s_{d-2}=\displaystyle\sum_{i=1}^dr_i^{d-2}=\frac{-a_{d-2}}{a_d}s_{d-4}=\cdots=(\frac{-a_{d-2}}{a_d})^{\frac{d-3}{2}}s_1=0
\end{equation}
So the sum of entropies vanishes, i.e. $\sum_i\mathcal{S}_i=0$.

In even dimensions,
\begin{eqnarray}
s_{d-2}=\displaystyle\sum_{i=1}^dr_i^{d-2}=\frac{-a_{d-2}}{a_d}s_{d-4}=\cdots=(\frac{-a_{d-2}}{a_d})^{\frac{d-2}{2}}s_0=2(\frac{(d-1)(d-2)k}{2\Lambda_f})^{\frac{d-2}{2}}.
\end{eqnarray}
So the entropy sum is
\begin{equation}
 \sum_i\mathcal{S}_i=\frac{\pi^{(d-1)/2}}{\Gamma(\frac{d-1}{2})}(1+f^{'}(R_0))(\frac{(d-1)(d-2)k}{2\Lambda_f})^{(d-2)/2} ,
\end{equation}
which does not depend on the mass $M$ and the conserved charge $Q$.

{\section{Conclusion and discussion}}
In order to investigate the property of entropy sum in all dimensions, we find that the formula \eqref{e5} is very useful for the calculation. By studying the maximally symmetric black holes in Lovelock gravity and $f(R)$ gravity and Kerr-(anti)de-Sitter black holes in Einstein gravity, we prove that the sum of all horizons indeed only depends on the coupling constants of the theory and the topology of the black hole, and does not depend on the conserved charges like $J_i$, $Q_i$ and mass $M$, therefore we can believe that it is a real universal property in all dimensions.
Especially, we have developed a method for calculating the entropy sum in the (even-dimensional) Kerr-(anti)de-Sitter case,
which can be used to calculate more complicated symmetric rational expressions
and may be useful for further study of universal entropy relations.

In this Letter, we have just discussed some special black hole solutions in several gravitational theories. It is important to verify this universal property in more general settings, i.e. black holes with less symmetry in more general gravitational theories with various matter contents. The rotating black holes in the Gauss-Bonnet (or even Lovelock) gravity are of special interest, whose exact analytical form for general parameters is not yet known. However, some approximate forms (e.g. in the slowly rotating case \cite{KC}) are known, which can be used to investigate the universal property of the entropy sum. The actual physics behind the universal properties that we have proved still needs more investigation. We wish to explore these aspects in future works.

{\section*{Acknowledgments}}
We thank Xiao-Ning Wu and Zhao-Yong Sun for useful discussions and comments. This work is supported by the Natural Science Foundation of China under Grant Nos. 11475179 and 11175245.

\end{document}